\documentclass[final]{vgtc}                          

\ifpdf
  \pdfoutput=1\relax                   
  \usepackage{graphicx}                
  \DeclareGraphicsExtensions{.pdf,.png,.jpg,.jpeg} 
\else
  \usepackage{graphicx}                
  \DeclareGraphicsExtensions{.eps}     
\fi%

\graphicspath{{figures/}{pictures/}{images/}{./}} 

\usepackage{microtype}                 
\PassOptionsToPackage{warn}{textcomp}  
\usepackage{textcomp}                  
\usepackage{mathptmx}                  
\usepackage{times}                     
\usepackage{cite}                      
\usepackage{tabu}                      
\usepackage{booktabs}                  

\usepackage{comment}

\onlineid{0}

\vgtccategory{Research}

\vgtcinsertpkg

\title{Design requirements to improve laparoscopy via XR}

\author{Ezequiel R. Zorzal\thanks{e-mail: ezorzal@unifesp.br}\\ %
     \parbox{2.5in}{\scriptsize \centering INESC-ID, Instituto Superior T\'ecnico \\ ICT/UNIFESP, Universidade Federal de S\~ao Paulo \\ Brazil} %
\and Maur\'icio~Sousa\thanks{e-mail: viva.o.mauricio@gmail.com}\\ %
     \scriptsize INESC-ID, University of Toronto \\ \scriptsize Canada %
\and Pedro~Belchior\thanks{e-mail: pedro.fls.belchior@gmail.com}\\ %
          \scriptsize INESC-ID, Instituto Superior T\'ecnico \\ \scriptsize Portugal \\%
\and João Madeiras Pereira\thanks{e-mail: jap@inesc-id.pt}\\ %
     \scriptsize INESC-ID, Instituto Superior T\'ecnico \\ \scriptsize Portugal %
\and Nuno~Figueiredo\thanks{e-mail: nuno.leitao.figueiredo@lusiadas.pt}\\ %
     \scriptsize Colorectal Surgery, Hospital Lusíadas, Lisbon \\ \scriptsize Portugal %
\and Joaquim~A.~Jorge\thanks{e-mail: jorgej@acm.org}\\ %
     \scriptsize INESC-ID, Instituto Superior T\'ecnico \\ \scriptsize Portugal %
     }

\abstract{ 
Laparoscopic surgery has the advantage of avoiding large open incisions and thereby decreasing blood loss, pain, and discomfort to patients. However, on the other side, it is hampered by restricted workspace, ambiguous communication, and surgeon fatigue caused by non-ergonomic head positioning. 
We aimed to identify critical problems and suggest design requirements and solutions.
We used user and task analysis methods to learn about practices performed in an operating room by observing surgeons in their working environment to understand how they performed tasks and achieved their intended goals. Drawing on observations and analysis from recorded laparoscopic surgeries, we have identified several constraints and design requirements to propose potential solutions to address the issues. 
Surgeons operate in a dimly lit environment, surrounded by monitors, and communicate through verbal commands and pointing gestures. Therefore, performing user and task analysis allowed us to understand the existing problems in laparoscopy better while identifying several communication constraints and design requirements, which a solution has to follow to address those problems.
Our contributions include identifying design requirements for laparoscopy surgery through a user and task analysis. These requirements propose design solutions towards improved surgeons' comfort and make the surgical procedure less laborious.%
} 

\CCScatlist{
  \CCScatTwelve{Human-centered computing}{Visu\-al\-iza\-tion}{Visualization design and evaluation methods}{}
}

\begin{document}




\firstsection{Introduction}

\maketitle


Laparoscopy - a minimally invasive surgical approach to the abdomen and pelvis - allows all types of resectional, staging and diagnostic procedures, especially when ultrasounds, CT scans and MRI scans cannot provide enough clinical information. Compared to open surgery, its major benefits are smaller and fewer incisions, with less tissue damage and pain following the operation, which in turn results in lower demands for analgesics and shorter hospital stays inducing a faster recovery time. However, Laparoscopic surgery 
affords diminished tactile feedback 
and also the loss of direct visual contact with organs, as surgeons rely on an endoscopic camera, which captures and feeds images onto a display~\cite{Bernhardt2016}. 
This results in a limited and restrictive experience compared to open surgery, as the surgeon dexterity and haptic feedback are reduced by the laparoscopic rigid instruments~\cite{Leite2016,Batmaz2017}. 

Furthermore, it presents other challenges that can induce surgeon discomfort by non-ergonomic positioning and hamper the surgical workflow. 
A paramount problem in minimally invasive surgery is hand-eye coordination, as surgeons have to look at screens placed outside the field of operation, which results in discomfort~\cite{Batmaz2017}, affecting the surgeon efficiency due to a visual-motor axis disconnection. One cannot look simultaneously at their instruments, hands and surgical field. 
To become proficient, high level training and experience is required to adapt to this condition, as extra mental effort must be applied~\cite{Leite2016}.
In addition, almost all these display screens are limited in sense that they do not support techniques to improve visual collaboration with the rest of the surgical team~\cite{HenryFuchs1MarkA.Livingston1RameshRaskar1DnardoColucci11963}.


Walczak et al.~\cite{Walczak2015} evaluated whether the positioning of the monitor has an impact on laparoscopic performance. 
Results show the time taken to perform the task was shorter when the screen was placed downwards, which corresponded to the position participants most preferred. This position allows users to flex the head at 15 to 45 degrees below eye level, which is the most comfortable position, as looking down improves eye lens accommodation and reduces eye weariness and headaches. 
\cite{Maithel2005} also evaluated the effect of the monitor, conducting a study to determine whether wearing a head-mounted display (HMD) improves task performance, or at least reduces muscle fatigue, comparing it in an operating scenario against the use of a traditional monitor. 
While the HMD 
made for smooth
 motion, but performance in general was not 
 clearly superior.

Batmaz et al.~\cite{Batmaz2017} compared 
visualisation, direct vision, 2D fish-eye, 
undistorted 
and 3D stereoscopic views, and studied their performance effects on a laparoscopic training exercise where 
subjects were to place a small object in the centre of five targets, in a specific order. 
The results show that 3D stereoscopic imaging does not have any performance edge over 2D, with objects being selectively coloured, facilitating depth perception. However, straight ahead monitor positioning did 
show performance benefits, as subjects took less time to perform the tasks as they felt less neck strain and more comfort. 
Prescher et al.~\cite{Prescherb} also assessed 
3D viewing, and conducted a study to determine whether 
stereoscopic 3D displays with glasses improved performance in trainees.
The 3D display 
seemingly reduced the time taken to complete the test as well as the number of dropped objects, while being generally preferred by 
subjects.

Kihara et al.~\cite{Kihara2012} developed a virtual reality (VR) system for use in real-world operation, combining a HMD with a 3D endoscope to provide the surgeon with high quality imaging right in front of him. The 3D HMD gives the feel of an open surgery and allows the visualization of content regardless of head position, while direct vision is allowed by lowering the angle of sight. Finally, 
\cite{Jayender} worked on a mixed reality headset which integrates the image from the laparoscopic camera, a navigation system and diagnostic imaging, complemented by an audio feedback system. 
VR can be successfully used for laparoscopic training curriculum as it not only helps to reduce the physical and mental workload on surgeons but also improves their surgical performance in operation theaters. It can also be used to train surgeons to cope up with other technical problems encountered during surgery simultaneously~\cite{Kantamaneni2021}.

Unlike previous work, 
we aim to identify potential design challenges in laparoscopic procedures and 
suggest design requirements 
to approach these problems. 
In this paper, we present a study on the conditions in which the laparoscopic surgeons perform their tasks. 
We analyze areas of improvement and contribute design requirements to 
appropach these issues. 


\section{Method}

We conduct our work according to the methodology used by~\cite{Mentis2014}. The examples presented in this paper 
come from the 
Champalimaud Foundation in Lisbon, PT~\cite{Zorzal2020}.
At that site, we 
observed laparoscopic surgeries on one of the Foundation's surgical rooms and 
talked to surgeons before and after the procedures, who explained 
what was about to happen or what took place. Also, during the surgery, nurses 
provided insight into the several stages of the surgery, or what was happening at that time. We also had the opportunity to ask questions of the surgeons at appropriate moments during the surgery. Besides that, we video recorded for further analysis in addition to our field notes. 
A total of five (5) laparoscopic surgeries were observed for a total of approximately 10 hours.

The observed cases included the surgical team is composed of mostly male members, with only one female surgeon in it, and their ages range from 34 to 42 years of age. 
During surgery, there are at least six people involved in the procedure:
A head surgeon, who coordinates the entire procedure, 1 or 2 assistant surgeons, who mostly observe but also participate in parts of the surgery, a nurse solely responsible for passing the surgeons tools they may require throughout the operation, an anesthetist keeping track of the patient's vital signs, a nurse supporting the anesthetist and a circulating nurse. Additionally, a senior surgeon may come in and serve as an advisor, providing insight and making remarks about what is being seen on camera.


\bgroup


We collected images and videos using smartphones and tablets, as well as the notes made during the observations. The inductive bottom-up approach to data analysis was used in which the authors analyzed their field notes and videos. The findings and the discussion are presented in the subsequent section. 

\section{Results and Discussion}

Performing user and task analysis allowed us to better understand the existing problems in the procedure of laparoscopy, while identifying several constraints and design requirements, which a solution has to follow in order to address those problems. For the following, we will discuss the design requirements identified through the analysis. Problems statements, requirements, and the proposed design solutions also are summarized in Table~\ref{table:problem_requirements_solution}.


\begin{table*}[htb]
\caption{Problem statements, design requirements and design solution for a prototype that supports laparoscopy}\label{table:problem_requirements_solution}
\begin{tabular}{|p{0.31\textwidth}|p{0.31\textwidth}|p{0.31\textwidth}|}
\hline
{\bfseries \small Problem Statements} & {\bfseries \small Design Requirements} & {\bfseries \small Design Solution}\\
\hline
\small Visualizing the laparoscopic video during extended periods of time is exhausting for the neck. & \small The solution should allow the user to adopt more comfortable neck postures instead of forcing the user to look to the side to see what the other surgeons are seeing. & \small Following display: The laparoscopic video follows user head movement, so users can look around and assume a neck posture that is more comfortable for them.  \\
\hline
    \small Current interactions surgeons have, such as pointing or consulting patient data, require them to let go of their tools, which interrupts the procedure. & \small Surgeons should have hands-free interactions in order to operate in an uninterrupted fashion. & \small Hands-free interaction: Every interaction is either done with the head or using the feet. \\
    \hline
    \small Browsing patient data interoperatively takes too long because it requires to call in an assistant, who browses the images for the surgeon. & \small Users should be able to look at patient data by themselves, without interrupting and adding extra time to the surgery. & \small Patient data image browser: users can look to the side to see and browse magnetic resonance images from the patient.  \\
    \hline
    \small Users may have to move around the patient in order to adopt better positions to hold their tools. & \small Interaction using the foot should not rely on pedals, as these would need to be moved around to cope with user movement. & \small Foot browsing: Users can use the foot to navigate the patient images, rotating it on its heel to change images faster or slower. \\
    \hline
    \small Pointing is unclear and ambiguous: different users have different interpretations of where a surgeon is pointing at. & \small Users should be able to point precisely and understand where other users are pointing at, regardless of position in the operating room. & \small Pointing reticle: users can place a reticle on both laparoscopic video and patient images, controlling it with head motion. This cursor is visible on other users’ headsets.\\
    \hline
    \small Surgeons operate in a crowded area, as they are usually very close together. & \small Augmented space should present information close to the surgeon to prevent it from appearing intersected with a colleague. & \small Close quarters: Positioning of interface elements is no further than at an elbow's reach.\\
    \hline
\end{tabular}
\end{table*}

\subsection{Following display}

Laparoscopy is an intensive process, not just mentally but physically as well. The procedure is already very demanding in itself due to surgeons having to expend extra mental effort thanks to a lack of hand-eye coordination that is caused by indirect visualization (Fig.~\ref{fig:neck}). That effort extends to the physical plane when we consider that they have look at the screen all the time, which places a continuous strain on their necks. Laparoscopy currently faces the glaring problem of monitor positioning.  During surgery, screens are usually placed far away and at an uncomfortable angle, causing neck and eye strain over the course of a surgery, especially if it drags for longer periods of time. Given this, it was important to allow the surgeons some freedom in how they want to see the video, and then the video,  while visible,  should follow user head movements so users do not have to reposition it in the augmented space, should they feel the need to assume another posture with the neck. 



\begin{figure*}[!htbp]
    \centering
    \includegraphics[width=0.65\textwidth,keepaspectratio]{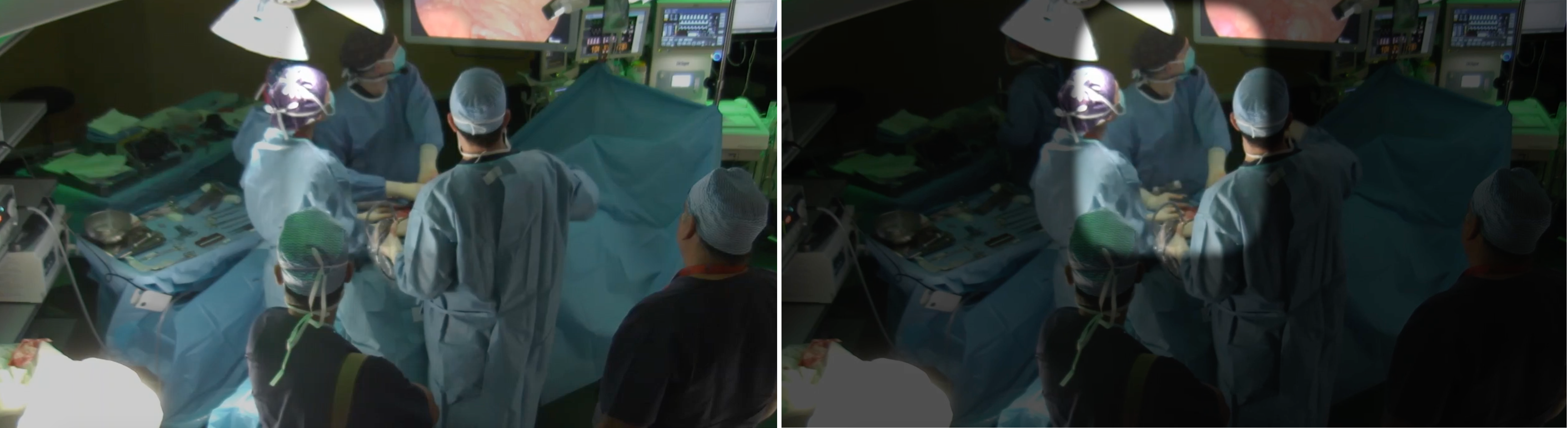}
    \caption{All doctors look to the same screen and, sometimes, this means having to assume an uncomfortable position. Visualizing the laparoscopic video during extended periods of time is exhausting for the neck.}
    \label{fig:neck}
\end{figure*} 

The surgeons have to look at screens placed outside the field of operation, which results in discomfort~\cite{Batmaz2017}, affecting the surgeon's efficiency due to a disconnect between the visual and motor axis, because the surgeon cannot look at the instruments or hands and the field of surgery simultaneously. To be successful, more training is required to adapt to this condition, as extra mental effort must be applied~\cite{Leite2016}.
In addition, almost all 
display screens are limited in sense that they do not support techniques to improve communication and visual collaboration with the rest of the surgical team~\cite{HenryFuchs1MarkA.Livingston1RameshRaskar1DnardoColucci11963,HMentis2019}.

Muratore et al.~\cite{Muratore2007}  suggest for the future, 
the ideal display system would be a 3D high definition image HMD, 
citing the comfort of looking at the endoscopic image in any preferred head position, improving ergonomics and reducing neck strain. The use of an HMD is also seen as beneficial in the sense that it alleviates equipment clutter in the operating room. It is further noted the usefulness of individualized image manipulation features like zooming, which allows each surgeon to see the endoscopic video in the way they find most comfortable. 
Also, the works of~\cite{Walczak2015,Maithel2005,Batmaz2017,Kihara2012} seem to support the usage of a HMD for laparoscopic surgery, with the video following the user’s head movements. With this, users can assume their preferred head position instead of being forced to look sideways in order to see the video.

\subsection{Hands-free interaction}

From our field observations, we noticed that surgeons place down their tools to perform some secondary tasks, which interrupts the procedure (Fig.~\ref{fig:losing}). We suggest that projects avoid this type of situation, with a completely hands-free approach, using both head gaze and foot movement as sources of input.


\begin{figure*}[!bhtpb]
    \centering
    \includegraphics[width=1\textwidth,keepaspectratio]{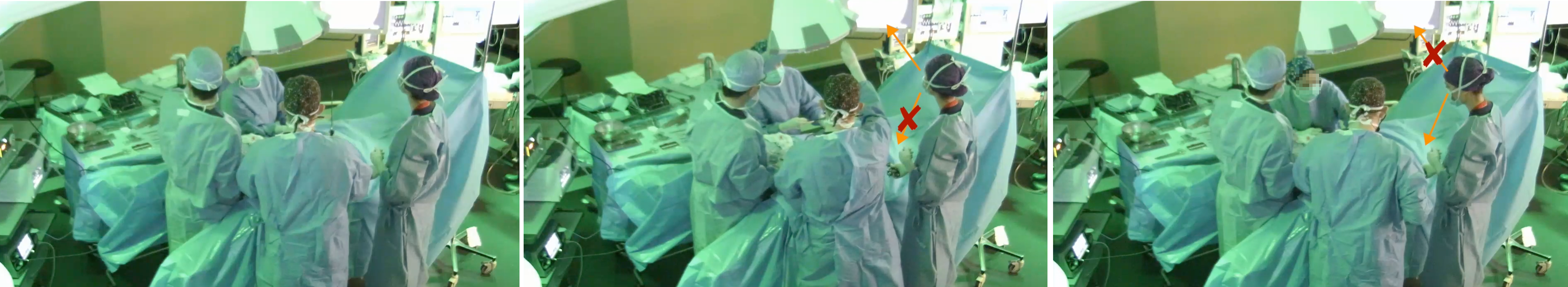}
    \caption{Surgeons can't look at their hands, thus losing hand-eye coordination.}
    \label{fig:losing}
\end{figure*}

Other studies~\cite{Kim2017,Jayender} 
looked at 
using head movements, and gaze 
to select targets, especially 
when the content follows head movements~\cite{Grinshpoon}. A more elaborate approach takes the form of the eye gazing~\cite{Esteves2015,Velloso2016}, which was well-received by users, but may not transition well onto the surgical operating field: these controls would have to be displayed continuously and right in front of the user, unlike in the presented works, which could be distractive for users, but more importantly, it would take valuable space from the HMD's already limited field of view. In terms of feet, two different approaches emerge: using a foot pedal as a means to activate a selected control~\cite{Jayender} and using foot movement to select and activate controls~\cite{Muller2019}. After comparing the two, we conclude using foot movement would be a more flexible choice, as it does not rely on extra hardware that is situated in a given position in space.

\subsection{Patient data image browser and foot browsing}


Consulting patient data intra-operatively, such as MRIs and computed tomographies, may be unfeasible. On the one hand, the data are extensive and may require some time to identify the required set. On the other, surgeons must abandon the operating table to sit at a computer and browse the desired images. The surgeon then gives directions on where to look and when to stop while an assistant handles the computer.  Surgeons usually do not browse the images themselves when sterilized. Indeed each interaction with non-sterile equipment such as the keyboard and mouse would entail a new sterilization procedure, wasting additional resources. This has been discussed in \cite{ohara-2014,Lopes-2019,Johnson2011}. This is a design opportunity for AR as discussed in \cite{Zorzal2019}.

In the work of 
\cite{Muratore2007}, the authors emphasize the importance of the HMD, stating preoperative imaging could be individually manipulated through its use, as well as grant surgeons extra comfort by allowing them to see the laparoscopic image regardless of head positioning. Hands-free interaction is again considered, with the suggestion of using foot pedals instead. Also discussed is the issue of paradoxical imaging, which occurs when the camera faces the surgeon, causing movements with the tool to appear inverted compared to the hand movements. However, in an interview with the surgeons of the 
Champalimaud Foundation,  this did not appear to be an issue, since the surgeon can move around the 
operating table, which ensures the camera always faces the opposite direction. This freedom to move around also impacts the practicality of using foot pedals to ensure hands-free interaction, as the authors suggest, as the surgeon would have to either have the same pedals on multiple sides, or move the pedals around. In this case, exploring foot movement, as proposed by 
\cite{Muller2019}, could be more useful.


\subsection{Pointing reticle and close quarters}

Although they are close quarters, surgeons also currently face problems in communication. In fact, according to the inquired surgeons, just as they complain about difficulty in maintaining proper posture, so do they complain about not being able to let other surgeons know what part of the video they are pointing at, or to understand what others are pointing at as well. Several works have approached this by looking at proxemics \cite{Mentis2012} and embodied vision~\cite{Mentis2013}. 



\begin{figure*}[!htpb]

    \centering
    \includegraphics[width=0.65\textwidth,keepaspectratio]{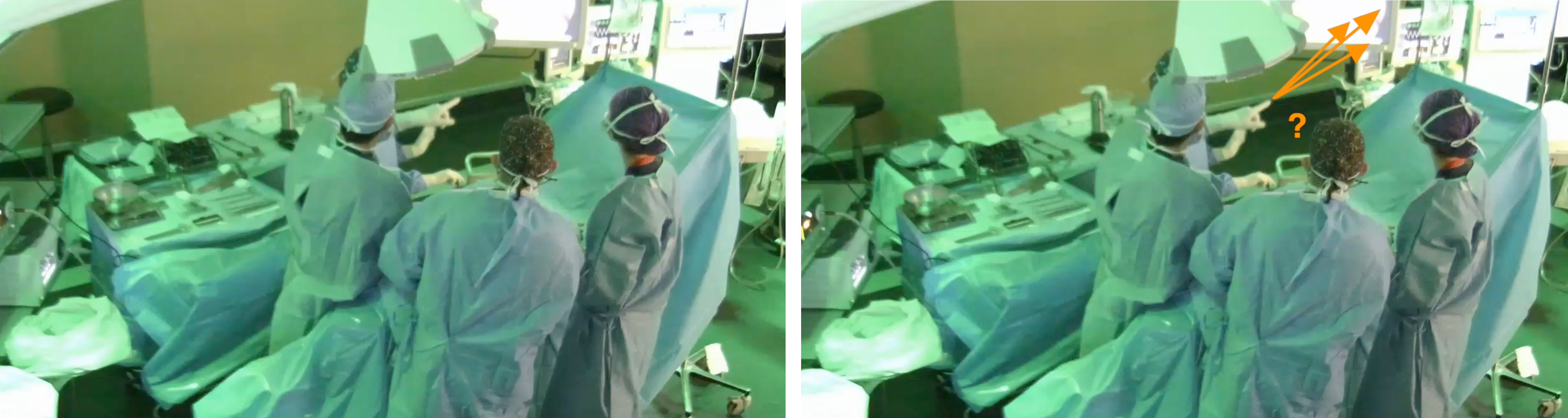}
    \caption{Doctor point at the screen to communicate. Communication is unclear and ambiguous: different users have different interpretations of where a surgeon is pointing at.}
    \label{fig:pointing2}
\end{figure*}


The instructor can point at the screen  
for the other surgeons to understand what anatomical structure he/she is referring to and 
use gestures for others to understand the motion of the tools better and envision cutting lines. Sometimes, pointing can also be done with the tools themselves, but even though it may be effective, it is not always correct because if both hands are occupied, it implies letting go of a structure to point with the tool or asking someone else to hold it. Additionally, pointing from a distance with the hand is ambiguous at best, as there is no clear way to tell where precisely a surgeon is pointing at, as can be seen in Fig.~\ref{fig:pointing2}.


For Prescher et al.~\cite{Prescher}, the impact of the pointer in a real operating scenario may be lessened because target selection is not random but rather contextual, meaning that the following targets may be located through the description of what is being displayed on-screen. Also, the pointer is embedded in the laparoscope. Thus the camera must be displaced to move the pointer, causing the view plane to change and forcing the surgeon to readjust to the new perspective, losing perceived depth. Therefore, it would be more useful if the cursor moved independently of the camera, controlled via gestures or head tracking for a hands-free approach, as suggested by 3D interactions performed above a table~\cite{mendes2014,mendes2016}.



\section{Conclusion}



This paper presents the requirements and design solutions for laparoscopy through a user and task analysis. During a task analysis, we map out the sequence of activities surgeons go through and the actions required to achieve that goal. Drawing on observations and analysis of video recordings of laparoscopic surgeries, we identify several constraints and design requirements, which a solution will have to follow in order to address those problems. These requirements propose to inform the design solutions towards improved surgeons' comfort and make the surgical procedure less laborious.

\section*{Acknowledgments}




This work was supported by national funds through FCT, Fundação para a Ciência e a Tecnologia, under project UIDB/50021/2020. The authors would like to thank the Champalimaud Foundation for its collaboration and support in the development of the study.







\bibliographystyle{abbrv-doi}

\bibliography{article}

\begin{thebibliography}{10}

\bibitem{Batmaz2017}
A.~U. Batmaz, M.~De~Mathelin, and B.~Dresp-Langley.
\newblock {Seeing virtual while acting real: Visual display and strategy
  effects on the time and precision of eye-hand coordination}.
\newblock {\em PLoS ONE}, 12(8):1--18, 2017. doi: {{%
10\hspace{.1pt}\discretionary{.}{%
}{.}\hspace{.4pt}1371\discretionary{/}{%
}{/}journal\hspace{.1pt}\discretionary{.}{%
}{.}\hspace{.4pt}pone\hspace{.1pt}\discretionary{.}{%
}{.}\hspace{.4pt}0183789}}


\bibitem{Bernhardt2016}
S.~Bernhardt, S.~A. Nicolau, V.~Agnus, L.~Soler, C.~Doignon, and J.~Marescaux.
\newblock {Automatic localization of endoscope in intraoperative CT image: A
  simple approach to augmented reality guidance in laparoscopic surgery}.
\newblock {\em Medical Image Analysis}, 30:130--143, 2016. doi: {{%
10\hspace{.1pt}\discretionary{.}{%
}{.}\hspace{.4pt}1016\discretionary{/}{%
}{/}j\hspace{.1pt}\discretionary{.}{%
}{.}\hspace{.4pt}media\hspace{.1pt}\discretionary{.}{%
}{.}\hspace{.4pt}2016\hspace{.1pt}\discretionary{.}{%
}{.}\hspace{.4pt}01\hspace{.1pt}\discretionary{.}{%
}{.}\hspace{.4pt}008}}


\bibitem{Esteves2015}
A.~Esteves, E.~Velloso, A.~Bulling, and H.~Gellersen.
\newblock {Orbits: Gaze Interaction for Smart Watches using Smooth Pursuit Eye
  Movements}.
\newblock {\em Proc. UIST '15}, (1):457--466, 2015. doi: {{%
10\hspace{.1pt}\discretionary{.}{%
}{.}\hspace{.4pt}1145\discretionary{/}{%
}{/}2807442\hspace{.1pt}\discretionary{.}{%
}{.}\hspace{.4pt}2807499}}


\bibitem{HMentis2019}
Y.~Feng, K.~Li, A.~Semsar, H.~McGowan, J.~Mun, H.~R. Zahiri, I.~George,
  A.~Park, A.~Kleinsmith, and H.~M. Mentis.
\newblock Communication cost of single-user gesturing tool in laparoscopic
  surgical training.
\newblock In {\em Proceedings of the 2019 CHI Conference on Human Factors in
  Computing Systems}, CHI ’19. ACM, New York, NY, USA, 2019. doi: {{%
10\hspace{.1pt}\discretionary{.}{%
}{.}\hspace{.4pt}1145\discretionary{/}{%
}{/}3290605\hspace{.1pt}\discretionary{.}{%
}{.}\hspace{.4pt}3300841}}


\bibitem{HenryFuchs1MarkA.Livingston1RameshRaskar1DnardoColucci11963}
H.~Fuchs, M.~A. Livingston, R.~Raskar, et~al.
\newblock {Augmented Reality Visualization for Laparoscopic Surgery}.
\newblock {\em Federation bulletin / Federation of State Medical Boards of the
  United States}, 50:316--322, 2006. doi: {{%
10\hspace{.1pt}\discretionary{.}{%
}{.}\hspace{.4pt}1007\discretionary{/}{%
}{/}BFb0056282}}


\bibitem{Grinshpoon}
A.~Grinshpoon, S.~Sadri, G.~J. Loeb, C.~Elvezio, S.~Siu, and S.~K. Feiner.
\newblock Hands-free augmented reality for vascular interventions.
\newblock In {\em ACM SIGGRAPH 2018 Emerging Technologies}, SIGGRAPH '18.
  ACMGrinshpoon, New York, NY, USA, 2018. doi: {{%
10\hspace{.1pt}\discretionary{.}{%
}{.}\hspace{.4pt}1145\discretionary{/}{%
}{/}3214907\hspace{.1pt}\discretionary{.}{%
}{.}\hspace{.4pt}3236462}}


\bibitem{Jayender}
J.~Jayender, B.~Xavier, F.~King, et~al.
\newblock {\em {A Novel Mixed Reality Navigation System for Laparoscopy
  Surgery}}, vol. 2878.
\newblock Springer International Publishing, 2018. doi: {{%
10\hspace{.1pt}\discretionary{.}{%
}{.}\hspace{.4pt}1007\discretionary{/}{%
}{/}b93811}}


\bibitem{Johnson2011}
R.~Johnson, K.~O'Hara, et~al.
\newblock Exploring the potential for touchless interaction in image-guided
  interventional radiology.
\newblock In {\em Proceedings of the SIGCHI Conference on Human Factors in
  Computing Systems}, CHI '11, p. 3323–3332. Association for Computing
  Machinery, New York, NY, USA, 2011. doi: {{%
10\hspace{.1pt}\discretionary{.}{%
}{.}\hspace{.4pt}1145\discretionary{/}{%
}{/}1978942\hspace{.1pt}\discretionary{.}{%
}{.}\hspace{.4pt}1979436}}


\bibitem{Kantamaneni2021}
K.~Kantamaneni, K.~Jalla, et~al.
\newblock Virtual reality as an affirmative spin-off to laparoscopic training:
  An updated review.
\newblock {\em Cureus}, Aug. 2021. doi: {{%
10\hspace{.1pt}\discretionary{.}{%
}{.}\hspace{.4pt}7759\discretionary{/}{%
}{/}cureus\hspace{.1pt}\discretionary{.}{%
}{.}\hspace{.4pt}17239}}


\bibitem{Kihara2012}
K.~Kihara, Y.~Fujii, H.~Masuda, et~al.
\newblock {New three-dimensional head-mounted display system, TMDU-S-3D system,
  for minimally invasive surgery application: Procedures for gasless
  single-port radical nephrectomy}.
\newblock {\em International Journal of Urology}, 19(9):886--889, 2012. doi:
  {{%
10\hspace{.1pt}\discretionary{.}{%
}{.}\hspace{.4pt}1111\discretionary{/}{%
}{/}j\hspace{.1pt}\discretionary{.}{%
}{.}\hspace{.4pt}1442\discretionary{%
}{-}{-}2042\hspace{.1pt}\discretionary{.}{%
}{.}\hspace{.4pt}2012\hspace{.1pt}\discretionary{.}{%
}{.}\hspace{.4pt}03044\hspace{.1pt}\discretionary{.}{%
}{.}\hspace{.4pt}x}}


\bibitem{Kim2017}
T.~Kim, B.~Saket, A.~Endert, and B.~MacIntyre.
\newblock {VisAR: Bringing Interactivity to Static Data Visualizations through
  Augmented Reality}.
\newblock 2017.

\bibitem{Leite2016}
M.~Leite, A.~F. Carvalho, P.~Costa, et~al.
\newblock {Assessment of laparoscopic skills performance: 2D versus 3D vision
  and classic instrument versus new hand-held robotic device for laparoscopy}.
\newblock {\em Surgical Innovation}, 23(1):52--61, 2016. doi: {{%
10\hspace{.1pt}\discretionary{.}{%
}{.}\hspace{.4pt}1177\discretionary{/}{%
}{/}1553350615585638}}


\bibitem{Lopes-2019}
D.~Lopes, F.~Relvas, S.~Paulo, Y.~Rekik, L.~Grisoni, and J.~Jorge.
\newblock Feetiche: Feet input for contactless hand gesture interaction.
\newblock In {\em The 17th International Conference on Virtual-Reality
  Continuum and Its Applications in Industry}, VRCAI '19. ACM, New York, NY,
  USA, 2019. doi: {{%
10\hspace{.1pt}\discretionary{.}{%
}{.}\hspace{.4pt}1145\discretionary{/}{%
}{/}3359997\hspace{.1pt}\discretionary{.}{%
}{.}\hspace{.4pt}3365704}}


\bibitem{Maithel2005}
S.~K. Maithel, L.~Villegas, N.~Stylopoulos, S.~Dawson, and D.~B. Jones.
\newblock {Simulated laparoscopy using a head-mounted display vs traditional
  video monitor: An assessment of performance and muscle fatigue}.
\newblock {\em Surgical Endoscopy and Other Interventional Techniques}, 2005.
  doi: {{%
10\hspace{.1pt}\discretionary{.}{%
}{.}\hspace{.4pt}1007\discretionary{/}{%
}{/}s00464\discretionary{%
}{-}{-}004\discretionary{%
}{-}{-}8177\discretionary{%
}{-}{-}6}}


\bibitem{mendes2014}
D.~Mendes, F.~Fonseca, B.~Araùjo, A.~Ferreira, and J.~Jorge.
\newblock Mid-air interactions above stereoscopic interactive tables.
\newblock In {\em 2014 IEEE Symposium on 3D User Interfaces (3DUI)}, pp. 3--10,
  2014. doi: {{%
10\hspace{.1pt}\discretionary{.}{%
}{.}\hspace{.4pt}1109\discretionary{/}{%
}{/}3DUI\hspace{.1pt}\discretionary{.}{%
}{.}\hspace{.4pt}2014\hspace{.1pt}\discretionary{.}{%
}{.}\hspace{.4pt}6798833}}


\bibitem{mendes2016}
D.~Mendes, F.~Relvas, A.~Ferreira, and J.~Jorge.
\newblock The benefits of dof separation in mid-air 3d object manipulation.
\newblock In {\em Proceedings of the 22nd ACM Conference on Virtual Reality
  Software and Technology}, VRST '16, p. 261–268. Association for Computing
  Machinery, New York, NY, USA, 2016. doi: {{%
10\hspace{.1pt}\discretionary{.}{%
}{.}\hspace{.4pt}1145\discretionary{/}{%
}{/}2993369\hspace{.1pt}\discretionary{.}{%
}{.}\hspace{.4pt}2993396}}


\bibitem{Mentis2014}
H.~M. Mentis, A.~Chellali, and S.~Schwaitzberg.
\newblock Learning to see the body: Supporting instructional practices in
  laparoscopic surgical procedures.
\newblock In {\em Proceedings of the SIGCHI Conference on Human Factors in
  Computing Systems}, CHI ’14, p. 2113–2122. ACM, New York, NY, USA, 2014.
  doi: {{%
10\hspace{.1pt}\discretionary{.}{%
}{.}\hspace{.4pt}1145\discretionary{/}{%
}{/}2556288\hspace{.1pt}\discretionary{.}{%
}{.}\hspace{.4pt}2557387}}


\bibitem{Mentis2012}
H.~M. Mentis, K.~O'Hara, A.~Sellen, and R.~Trivedi.
\newblock Interaction proxemics and image use in neurosurgery.
\newblock In {\em Proceedings of the SIGCHI Conference on Human Factors in
  Computing Systems}, CHI '12, p. 927–936. Association for Computing
  Machinery, New York, NY, USA, 2012. doi: {{%
10\hspace{.1pt}\discretionary{.}{%
}{.}\hspace{.4pt}1145\discretionary{/}{%
}{/}2207676\hspace{.1pt}\discretionary{.}{%
}{.}\hspace{.4pt}2208536}}


\bibitem{Mentis2013}
H.~M. Mentis and A.~S. Taylor.
\newblock Imaging the body: Embodied vision in minimally invasive surgery.
\newblock In {\em Proceedings of the SIGCHI Conference on Human Factors in
  Computing Systems}, CHI '13, p. 1479–1488. Association for Computing
  Machinery, New York, NY, USA, 2013. doi: {{%
10\hspace{.1pt}\discretionary{.}{%
}{.}\hspace{.4pt}1145\discretionary{/}{%
}{/}2470654\hspace{.1pt}\discretionary{.}{%
}{.}\hspace{.4pt}2466197}}


\bibitem{Muller2019}
F.~M{\"{u}}ller, J.~McManus, S.~G{\"{u}}nther, M.~Schmitz,
  M.~M{\"{u}}hlh{\"{a}}user, and M.~Funk.
\newblock {Mind the Tap}.
\newblock In {\em Proceedings of the 2019 CHI Conference on Human Factors in
  Computing Systems - CHI '19}, pp. 1--13, 2019. doi: {{%
10\hspace{.1pt}\discretionary{.}{%
}{.}\hspace{.4pt}1145\discretionary{/}{%
}{/}3290605\hspace{.1pt}\discretionary{.}{%
}{.}\hspace{.4pt}3300707}}


\bibitem{Muratore2007}
C.~S. Muratore, B.~A. Ryder, and F.~I. Luks.
\newblock {Image display in endoscopic surgery History of surgical endoscopy}.
\newblock pp. 349--356, 2007.

\bibitem{ohara-2014}
K.~O'Hara, G.~Gonzalez, et~al.
\newblock {Touchless interaction in surgery}.
\newblock {\em Communications of the ACM}, 57(1):70--77, 2014. doi: {{%
10\hspace{.1pt}\discretionary{.}{%
}{.}\hspace{.4pt}1145\discretionary{/}{%
}{/}2541883\hspace{.1pt}\discretionary{.}{%
}{.}\hspace{.4pt}2541899}}


\bibitem{Prescher}
H.~Prescher, D.~E. Biffar, C.~A. Galvani, J.~W. Rozenblit, and A.~J. Hamilton.
\newblock {Surgical navigation pointer facilitates identification of targets in
  a simulated environment}.
\newblock In {\em Simulation Series}, vol.~46, pp. 246--252, 2014.

\bibitem{Prescherb}
H.~Prescher, D.~E. Biffar, J.~Rozenblit, and A.~J. Hamilton.
\newblock {The comparison of high definition versus stereoscopic display on
  standardized fundamental laparoscopic skill procedures}.
\newblock {\em Summer Computer Simulation Conference, SCSC 2014, Part of the
  2014 Summer Simulation Multiconference, SummerSim 2014, July 6, 2014 - July
  10}, 46(10):346--351, 2014.

\bibitem{Velloso2016}
E.~Velloso, M.~Wirth, et~al.
\newblock {AmbiGaze: Direct Control of Ambient Devices by Gaze}, 2016. doi: {{%
10\hspace{.1pt}\discretionary{.}{%
}{.}\hspace{.4pt}1145\discretionary{/}{%
}{/}2901790\hspace{.1pt}\discretionary{.}{%
}{.}\hspace{.4pt}2901867}}


\bibitem{Walczak2015}
D.~A. Walczak, D.~Pawe{\l}czak, P.~Piotrowski, et~al.
\newblock {Video display during laparoscopy – where should it be placed?}
\newblock {\em Videosurgery and Other Miniinvasive Techniques}, 1:87--91, 2015.
  doi: {{%
10\hspace{.1pt}\discretionary{.}{%
}{.}\hspace{.4pt}5114\discretionary{/}{%
}{/}wiitm\hspace{.1pt}\discretionary{.}{%
}{.}\hspace{.4pt}2014\hspace{.1pt}\discretionary{.}{%
}{.}\hspace{.4pt}47434}}


\bibitem{Zorzal2020}
E.~R. Zorzal, J.~M.~C. Gomes, M.~Sousa, P.~Belchior, P.~G. da~Silva,
  N.~Figueiredo, D.~S. Lopes, and J.~Jorge.
\newblock Laparoscopy with augmented reality adaptations.
\newblock {\em Journal of Biomedical Informatics}, 107:103463, July 2020. doi:
  {{%
10\hspace{.1pt}\discretionary{.}{%
}{.}\hspace{.4pt}1016\discretionary{/}{%
}{/}j\hspace{.1pt}\discretionary{.}{%
}{.}\hspace{.4pt}jbi\hspace{.1pt}\discretionary{.}{%
}{.}\hspace{.4pt}2020\hspace{.1pt}\discretionary{.}{%
}{.}\hspace{.4pt}103463}}


\bibitem{Zorzal2019}
E.~R. Zorzal, M.~Sousa, D.~Mendes, R.~K. dos Anjos, D.~Medeiros, S.~F. Paulo,
  P.~Rodrigues, J.~J. Mendes, V.~Delmas, J.-F. Uhl, J.~Mogorrón, J.~A. Jorge,
  and D.~S. Lopes.
\newblock Anatomy studio: A tool for virtual dissection through augmented 3d
  reconstruction.
\newblock {\em Computers \& Graphics}, 85:74 -- 84, 2019. doi: {{%
10\hspace{.1pt}\discretionary{.}{%
}{.}\hspace{.4pt}1016\discretionary{/}{%
}{/}j\hspace{.1pt}\discretionary{.}{%
}{.}\hspace{.4pt}cag\hspace{.1pt}\discretionary{.}{%
}{.}\hspace{.4pt}2019\hspace{.1pt}\discretionary{.}{%
}{.}\hspace{.4pt}09\hspace{.1pt}\discretionary{.}{%
}{.}\hspace{.4pt}006}}


\end{thebibliography}

\end{document}